\documentclass[twocolumn]{aastex63}

\def\lapp{\ifmmode\stackrel{<}{_{\sim}}\else$\stackrel{<}{_{\sim}}$\fi}
\def\gapp{\ifmmode\stackrel{>}{_{\sim}}\else$\stackrel{>}{_{\sim}}$\fi}
\usepackage{multirow}
\usepackage{color}
\usepackage{amsmath}
\usepackage{soul}
\usepackage{gensymb}
\usepackage{hyperref}
\usepackage{lineno}

\accepted{May 26, 2020}
\submitjournal{ApJL}

\begin{document}

\shorttitle{PSR~J2030+4415's PWN \& Filament}
\shortauthors{}

\title{PSR~J2030+4415's Remarkable Bow Shock, PWN and Filament}

\correspondingauthor{Martijn de Vries}
\email{mndvries@stanford.edu}

\author{Martijn de Vries}
\affiliation{Department of Physics/KIPAC, 
Stanford University, Stanford, CA 94305-4060, USA}
\author{Roger W. Romani}
\affiliation{Department of Physics/KIPAC, 
Stanford University, Stanford, CA 94305-4060, USA}

\begin{abstract}
We report on new X-ray and optical observations of PSR~J2030+4415, a Gamma-ray Pulsar with an H$\alpha$ bow shock. These data reveal the velocity structure of the bow shock apex and resolve unusual X-ray structure in its interior. In addition the system displays a very long, thin filament, extending at least $5^\prime$ at $\sim 130^\circ$ to the pulsar motion vector. Careful astrometry, compared with a short archival exposure, detects the pulsar proper motion at 85\,mas\,yr$^{-1}$. With the H$\alpha$ velocity structure this allows us to estimate the distance as $0.75$\,kpc.
\end{abstract}


\section{Introduction} \label{sec:intro}

J2030 is a $P=308$\,ms, $\tau = 6\times 10^5$\,yr radio-quiet pulsar discovered via $\gamma$-ray pulsations \citep{Pletsch2012}. With ${\dot E} = 2.2 \times 10^{34}{\rm\,erg\,s^{-1}}$, the heuristic $L_\gamma = ({\dot E} \cdot 10^{34} {\rm\, erg/s})^{1/2}$ law estimates $d \sim 0.8$\,kpc; no DM distance is possible. However, this is one of a handful of pulsars sporting an H$\alpha$ bow shock nebula \citep[][hereafter BR14]{Brownsberger2014}, This line emission lets us probe the pulsar distance and kinematics. In BR14 we noted a SWIFT X-ray detection, as expected for a nearby, high-${\dot E}$ PSR. A confirming 25\,ks ACIS exposure \citep[ObsID 14827, briefly described in][]{Marelli2015} separates the PSR from a faint $\sim 25^{\prime\prime}$ PWN trail.

\begin{figure}
\hspace*{-5mm}\includegraphics[scale=0.49]{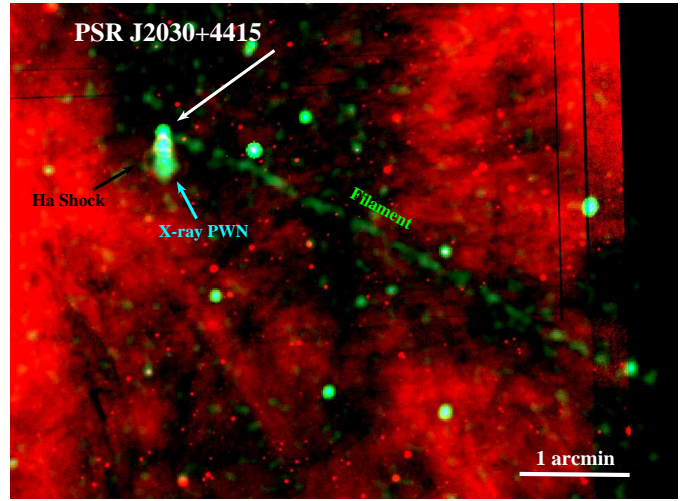}
\caption{The PSR J2030+4415 field with the adaptively smoothed 0.5-5\,keV {\it CXO} photons (shown in green) superposed on a narrow band H$\alpha$ image (shown in red). The faint thin filament extends to the edge of the ACIS-S field of view at right.}
\label{fig:HaX}
\end{figure}

The BR14 H$\alpha$ image showed two bubbles, trailing the pulsar. A Gemini GMOS IFU exposure, described below, finds a fainter H$\alpha$ bubble at the apex, surrounding the pulsar. The PWN trail extends through all three bubbles. We obtained a new, deeper {\it CXO} exposure to explore the relationship of the reverse (PWN X-ray) shock with the forward (H$\alpha$) bow shock. These data measure a proper motion shift of the X-ray source coincident with the pulsar, reveal complex structure in the PWN interior and discover a remarkable narrow pulsar filament extending at large angle to the proper motion axis (Figure \ref{fig:HaX}). 

\section{Observations} \label{sec:sec2}

\begin{figure*}
\centering
\includegraphics[scale=0.99]{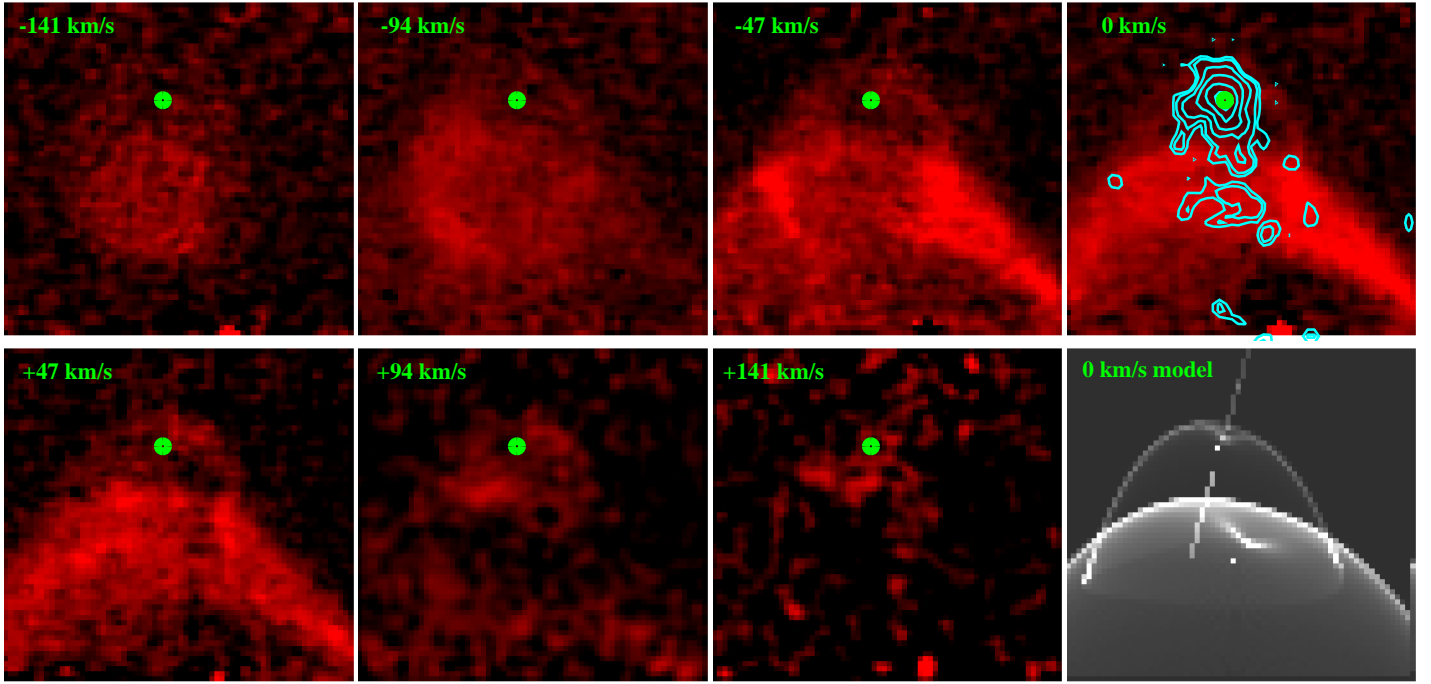}
\caption{\footnotesize Selected H$\alpha$ radial velocity slices from the GMOS-N IFU data cube. Frames span $6.5^{\prime\prime} \times 6.8^{\prime\prime}$ in steps of 47\,km s$^{-1}$. The pulsar position is marked by a green dot. The $\sim 0.5^{\prime\prime}$ standoff to the bubble 1 leading edge is best visible in the -47km/s to +47km/s frames, while the velocity extrema help pin down the spin (polar outflow) axis. Cyan contours on the $RV=0$ km/s frame show the {\it CXO} emission, adjusted for pulsar motion to the H$\alpha$ epoch. The zero-velocity slice of a \citet{Wilkin2000}-style model for bubbles 1 and 2 is shown in the last panel, with the pulsar spin axis marked with a line. Compare with Figs. \ref{fig:HaX} and \ref{fig:pm} for the larger scale X-ray emission.}
\label{fig:PWN}
\end{figure*}

\subsection{H$\alpha$ bow shock Measurements}

We observed J2030 on June 23, 2015 with GMOS-N using the R831 grating and the IFU, covering $5^{\prime\prime} \times 7^{\prime\prime}$, sampled at $0.2^{\prime\prime}$, with 0.339\,\AA/pixel spectral resolution  (Program GN-2015-FT17). J2030 was observed $5\times 1200$\,s, along with standard calibrations. Conditions were excellent with $\sim 0.35^{\prime\prime}$ FWHM measured in both an $r$ acquisition image and in the final data cube, which was assembled from a dithered combination of the exposures, using the GMOS IRAF analysis package. A single field star near the PWN apex was detected in continuum, verifying the registration and astrometry. Figure \ref{fig:PWN} shows a selection of the velocity channel images, which resolve the H$\alpha$ structure of the bow shock in 15.5\,km s$^{-1}$ steps. For reference, the X-ray brightness distribution from the pulsar and PWN apex are shown by cyan contours in the zero-velocity panel.

Pulsar bow shocks are non-radiative, and the shock limb shows up best in the low velocity channels dominated by H$\alpha$ from charge transfer from neutrals at the external medium velocity drifting into the shock. Evidently the pulsar is embedded in a faint apex bubble, which appears like a classical bow shock in these low velocity channels. At low velocity we also see the bow shock-like structure of the second bubble terminating behind the pulsar. The higher velocity channels mostly show the apex bubble, which appears to be driving outflow at $\pm 150$\,km s$^{-1}$. This outflow is markedly asymmetric with the negative (blue shifted) velocity channels peaking south of the pulsar. This suggests a somewhat elliptical cross section for this bubble with the major axis being driven, plausibly by pulsar polar outflow, at an angle of $\sim 30^\circ$ to the line of sight. We can model this H$\alpha$ structure following the simple bow shock computations of \citet{Wilkin2000}. The zero velocity channel of a model with bow shocks for bubbles 1 and 2 and a velocity mis-aligned pulsar spin axis is shown in the final panel of Figure \ref{fig:PWN}. Alas, the model does not follow the detailed H$\alpha$ structure closely enough for a direct fit.

\subsection{{\it CXO} X-ray Measurements}

We observed J2030 with 4 \textit{CXO} exposures on April 10-15, 2019 [ObsIDs 20298 (45\,ks), 22171 (40\,ks), 22172 (45\,ks), 22173 (23\,ks); $\sim 153$\,ks total] and compared with an archival exposure [ObsIDs 14827 (42\,ks)] from April 15, 2014. All exposures placed the target on the ACIS-S S3 chip and exposed in VF mode. The data were processed and analysed using \textit{CIAO} 4.12 and CALDB 4.9.1. 

\begin{figure}
\vspace{-28mm}
\centering
\hspace*{-5mm}\includegraphics[width=3.7 in]{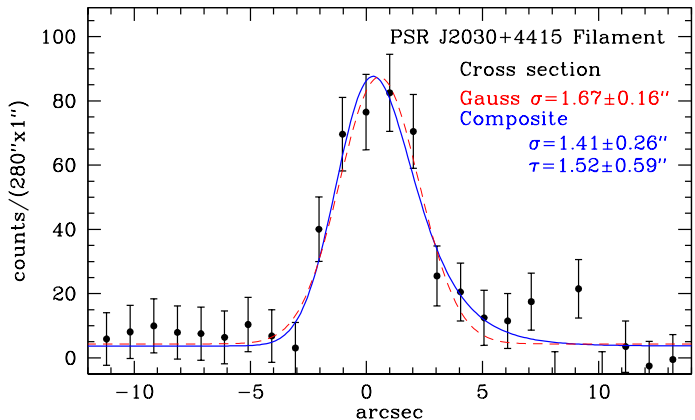} \\
\figcaption{Filament cross section (abscissa opposes the proper motion direction) averaged along its length. The dashed curve is a fit with a simple Gaussian, the solid curve show a composite model, fit as a falling exponential convolved with a Gaussian.
\label{fig:sbcut}
}
\vspace{-1mm}
\end{figure}

An unresolved X-ray source coincident with the pulsar, and the PWN trail are well visible in the new image. In addition we have discovered a remarkable, narrow X-ray filament extending for $5^\prime$ to the edge of the S3 chip, at $70^\circ$ to the PWN symmetry axis. The filament is narrow (Figure \ref{fig:sbcut}); if fit with a simple Gaussian we get a width $\sigma = 1.67\pm 0.16^{\prime\prime}$.

\begin{figure}
\centering
\vspace{-2mm}
\hspace*{-5mm}\includegraphics[width=3.7 in]{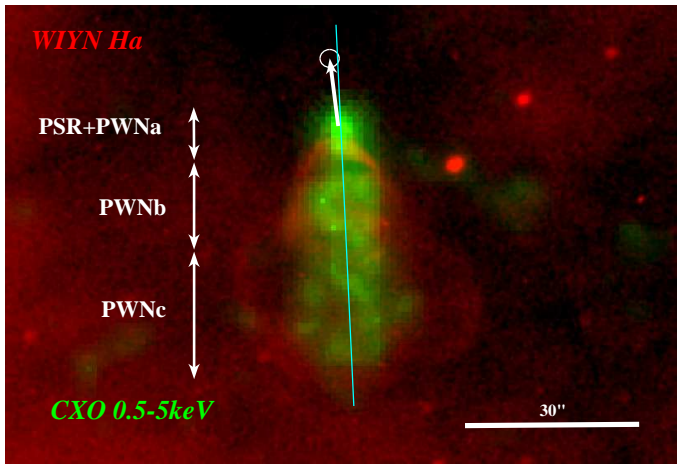} \\
\figcaption{Close-up of the PWN, showing the vertical range of the X-ray regions for the three bubbles (note Bubble 1 is not visible in this H$\alpha$ stretch -- see Figure \ref{fig:PWN}), the symmetry axis and the proper motion position, extrapolated forward 100\,yr.
\label{fig:pm}
}
\vspace{-2mm}
\end{figure}

The PWN itself also has an unusual structure, with the X-ray counts trailing in a wedge behind the pulsar through bubble 1, filling bubble 2 with a hollow cavity, and then continuing at near constant width through bubble 3. An adaptively smoothed X-ray image gives the impression of spiral or braided filamentary structure (Figure \ref{fig:pm}).

\medskip
\leftline{\bf X-ray proper motion}
\medskip

The H$\alpha$ velocity range implies $v_\perp \approx 300$km s$^{-1}$. We have used a detailed astrometric measurement of the point sources in this image, compared with the 2014 exposure to directly measure the pulsar proper motion. The basic technique compares the observed counts to the Poisson probability of PSF models generated for the spectrum and position of the point sources in each frame. This `Figure of Merit' (FoM) method was used in \citet{Etten2012} and \citet{Auchettl2015}, and is further developed here to push the frame registration and pulsar motion accuracy as close as possible to the statistical limit imposed by total counts for a collection of reference sources. We summarize the steps of this method below, for a more detailed discussion of the FoM technique see de Vries et al., in prep. 

To register the 2019 frames to the 2014 reference, we selected 16 point sources on the S2 and S3 chips, using the CIAO tool \textit{wavdetect} to obtain their positions. For each source we extracted the spectrum and produced a simulated PSF model at its reference position in each frame, using \textsc{marx}, with a simulated 5\,Ms exposure. We then generated images of each source and its PSF model, centered on the reference position, using 1/8th ACIS pixel bins, restricting to 0.5-3.5\,keV to minimize particle background and high energy PSF broadening. With these images, we constructed the FoM array, shifting the PSF model along the $x$ and $y$ axes. At each step, we calculated the Poisson probability of the data given the PSF model, using all pixels within a $3.5$ pixel circular aperture. The offset between the data and the PSF model, and thus the source-reference position offset, can then be determined from the centroid of a 2D Gaussian fit to the FoM array.

FoM fits were first used to correct the reference frame \textit{wavdetect} source positions. For the 2019 frames, these positions were offset with the measured GAIA proper motions of the optical counterparts, when available. For each frame we then summed all the data and model images, and constructed a single, coherent FoM to determine the frame offset. The statistical uncertainties in each frame are determined as $\sim \theta_{\rm PSF} /N^{1/2}$, and range from $20$ to $33$\,mas. Additionally, we estimate a systematic uncertainty of $40$\,mas by varying the PSF energy cuts and aperture sizes.

The pulsar positions in each frame were also determined using the FoM technique. When constructing the FoM, we masked the PWN, ignoring those pixels in the fit to minimize PWN influence on the measured PSR position. We used the same energy range as for the determining the frame shifts (0.5-3.5\,keV), and a slightly smaller 3 pixel radius aperture. With this single source the positional statistical uncertainty of an exposure is $37-63$\,mas. We estimate a systematic uncertainty of $30$\,mas. The total uncertainty for the proper motion in each frame is then determined by adding in quadrature all statistical and systematic uncertainties for both the frame offset and the pulsar position. 

Averaged over the four 2019 frames, we find a proper motion $\mu_{\rm RA} = 15 \pm 11$\,mas\,yr$^{-1}$ and $\rm \mu_{\rm DEC} = 84 \pm 12$\,mas\,yr$^{-1}$, for a total motion of $85 \pm 16$\, mas\,yr$^{-1}$. We considered the effects of differential rotation around the Galactic centre, as well as the motion of the sun with respect to the local standard of rest. At $d = 0.8$ kpc and $l \approx 82 \degree$, Galactic differential rotation contributes negligibly to this proper motion. However to best estimate the pulsars motion in its local standard of rest, we did correct for the Solar motion \citep{Schoenrich2010}, which induces an apparent proper motion $\mu_{\rm l} \approx 2$ mas yr$^{-1}$ and $\mu_{\rm b} \approx -1.5$ mas yr$^{-1}$. Correcting for these effects shifts the true direction of J2030's motion slightly SW to $PA=7.2\pm 7.6^\circ$.

The final proper motion offset (extrapolating 100\,yr) and its uncertainty are shown in Figure \ref{fig:pm}. The best fit position angle is slightly East, but within 1$\sigma$ of the PWN symmetry axis; local ISM motions may produce some of this apparent offset. The amplitude indicates a transverse velocity $v = 404\pm 86 \,d_{\rm kpc}$\,km\,s$^{-1}$, with the IFU velocity spread indicating $\sim 300\,{\rm km\,s^{-1}}$ this gives $d_{\rm kpc} \approx 0.75$, supporting the heuristic $L_\gamma$ distance estimate.

\medskip
\leftline{\bf Spectral Analysis}
\medskip

The results of our basic spectral analysis of the PSR, PWN and filament regions are shown in Table \ref{table:fits}. The PSR counts were extracted using a small $1.5^{\prime\prime}$ radius aperture (applying an ARF correction with the CIAO tool \textit{arfcorr}) in order to minimize contamination from the PWN. The PWN and filament regions were each fit as whole objects and split into subregions: PWN regions a, b, and c correspond to the 3 H$\alpha$ bubbles (best observed in Figure \ref{fig:pm}), while Fil a and b cover the eastern and western halves of the filament respectively.

\citet{Marelli2015} fit the PSR spectrum as a simple power law (PL), but it more likely combines thermal surface emission with magnetospheric PL. We therefore fit the point source with a composite PL+BB model and use a PL for all other regions. The Galactic hydrogen column density $N_H$ is poorly constrained in the 2019 data. When left as a free parameter, we obtain $N_H \approx 4.5 \times 10^{21}$ cm$^{-2}$, an unrealistically large value for $d=0.8$ kpc at these coordinates. Because the 2019 data is not sufficiently constraining, we have fixed the nH value to two different values in our analysis: $N_H = 0.6 \times 10^{21}$ cm$^{-2}$, the value used by \citet{Marelli2015}, and $N_H = 2 \times 10^{21}$ cm$^{-2}$, within the errors of our spectral fits. For the low absorption, we find an equivalent blackbody radius of $0.17 d_{\rm kpc}$km, while the high $N_H$ gives $0.4 d_{\rm kpc}$km, so in either case this would represent a heated polar cap region. The PL luminosity is $L_{2-10{\rm keV}} \approx 1.2\times 10^{30} d_{\rm kpc}^2 {\rm erg\,s^{-1}}$, which would match the heuristic \citet{Possenti2002} ${\dot E}$ scaling at $\sim 2$\,kpc. However, given the poor separation of the thermal and PL components, all we can say is that the fluxes are comparable to those of other young pulsars.

\begin{table}[b!!]
\caption{Spectral fit results for PSR, PWN and filament} 
\smallskip
\centering 
\begin{tabular}{l l l l } 
\hline\hline 
Comp		& Counts & kT/$\Gamma_{2.0}$ &kT/$\Gamma_{0.6}$ \\[0.5ex]
\hline 
PSR kT     	&$181\pm14$ &$0.11\pm0.02$ &$0.13\pm0.03$ \\
PSR PL	 	& "         &$1.43 \pm0.39$  &$1.31\pm0.40$ \\
PWN    		&$396\pm33$ &$1.72\pm0.11$  &$1.48\pm0.10$ \\
PWN a & $65\pm8$ & $1.15\pm0.24$& $1.31\pm0.26$   \\
PWN b & $173\pm14$ & $1.39\pm0.15$ & $1.61\pm0.16$\\
PWN c & $136\pm13$ & $1.87\pm0.19$ & $2.17\pm0.21$\\
Fil   &$330\pm55$ &$1.41\pm0.16$  &$1.24\pm0.14$ \\
Fil a	&$203\pm40$ &$1.35\pm0.20$  &$1.18\pm0.20$ \\
Fil b	&$117\pm37$ &$1.49\pm0.25$  &$1.32\pm0.24$ \\
\hline
\end{tabular}
\label{table:fits} 
\leftline{$N_H$ fixed at $2.0$ or $0.6\times10^{21}{\rm cm^{-2}}$. kT units keV.}
\vskip -0.5truecm
\end{table}

Combined, the PWN components have a flux $f_{\rm 0.5-10 keV} \approx 3.7 \times 10^{-14} {\rm erg\,cm^{-2}\,s^{-1}}$ for a luminosity $L_{\rm PWN} = 4.4 \times 10^{30} {\rm erg\,s^{-1}}$, nearly independent of $N_H$. This is in agreement with the rough $L_{PWN} \approx 10^{-3}{\dot E}$ scaling \citep[e.g.][]{Possenti2002} at $d=0.75\,$kpc. Also independent of $N_H$ the PWN flow softens downstream in the PWN with $\Delta \Gamma \approx 0.8 \pm 0.3$ ($2.5\sigma$).  This suggests substantial $e+/e-$ cooling in the 300\,yr it took the pulsar to traverse this structure. In contrast, there is no significant evidence for cooling along the filament, suggesting fast flow -- this spectral conclusion is tentative, as a $\Delta \Gamma=0.5$ cooling break is allowed at the $2 \sigma$ level, but the near-constant surface brightness along the filament argues against substantial cooling. The $\tau \approx 300$\,yr cooling of the $\sim$keV PWN synchrotron spectrum indicates typically energetic $\gamma \approx 8 \times 10^7 E_{\rm keV}^{2/3} \tau_{300}^{1/3}$ $e^\pm$ and a modestly boosted post-shock PWN field $33 E_{\rm keV}^{-1/3} \tau_{300}^{-2/3}\mu$G.

\section{Filament Morphology and Conclusions}

Only a handful of pulsar filaments are known, the most famous associated with the Guitar nebula/PSR B2224+65 \citep{Bandiera2008}; others are the `Lighthouse'/PSR J1101$-$6101 \citep{Pavan2016} and PSR J1509$-$5058 \citep{Klingler2016}. As seen in the `Lighthouse', our filament flux is nearly identical to that of the PWN (albeit with a harder spectrum); for J1509$-$5058, the filament shows about half the PWN flux, for the Guitar, the PWN is not detected. These PSRs all feature high velocity and a dense upstream medium, ensuring a small stand-off distance for the bow shock and facilitating escape of TeV PSR/PWN $e^\pm$. J2030 bears many similarities to the Guitar system (with a multi-bubble H$\alpha$ bow shock).
 
The basic picture for pulsar filaments described by \citet{Bandiera2008} has been elaborated by later authors \citep[e.g.][]{Olmi2019, Barkov2019}. Their one-sided nature is posited to track escape to the ISM from a special set of PWN field lines, where the polar PSR/PWN field breaks the magnetospheric symmetry. The leading hemisphere's field sign determines which side of the bow shock injects most efficiently to the filament. Here it is intriguing that the H$\alpha$ line maps imply that the spin axis is misaligned with the velocity, pointing NNW into the plane of the sky. The escape must be strongly associated to the western side of the PWN, since we see no evidence for a counter-filament. Thus the forward-facing magnetosphere polarity opposes that of the ambient ISM field. Figure \ref{fig:HaX} shows striation patterns in the H$\alpha$ background, roughly parallel to the filament, suggesting that the ISM B-field aligns with this axis. 

The filament width is likely over-estimated, as with the limited counts it proved difficult to follow the slightly curved filament path; thus our length-averaged cross section likely suffers some artificial broadening. However assuming that the $\gamma_7 \approx 8$ PWN $e^\pm$ derived above illuminate the filament, confining the Larmor radius within its Gaussian width $\sigma =1.7^{\prime\prime}$ puts a weak lower bound on the filament $B$ field of $>5/d_{kpc}\mu$G. In the case of the Guitar, the filament clearly moves along with the pulsar and has a quasi-exponential fall-off behind. J2030's filament does seem to have some excess counts behind the leading edge (to the right in Figure \ref{fig:sbcut}). If we fit to a falling exponential convolved with a Gaussian, $\chi^2$ decreases by 1.5 from the simple Gaussian and we find a tail scale $\tau=1.52\pm0.59^{\prime\prime}$. However, both models are quite acceptable, with $\chi^2/DoF = 0.82$ (Gaussian) and 0.80 (composite model).

Interpreting the $\tau_c \approx 1.5^{\prime\prime}$ exponential tail as a fading residual of a moving filament implies a cooling time $\tau_c/\mu \approx 18$\,y, independent of distance. This implies a much larger typical filament field $B \sim 100 \mu$G (if the cooling is faster, $B$ increases as $\tau_c^{-2/3}$). This substantially exceeds typical ISM fields, but may be generated by the injected particles. Even more interesting is the lack of cooling, or even surface brightness decrease, along the $l = 5^\prime$ to the edge of the field of view. To avoid cooling, the flow {\it along} the field lines must be very rapid, $\approx l v_{PSR}/\tau_c \approx 6 \times 10^4 {\rm\,km\, s^{-1}}$ or $\sim c/5$. Thus, the particles nearly free-stream along the filament. It would be interesting to trace the filament's full extent and eventual fading, as this probes the long-field propagation -- a timely subject given current interest in $e^+$ escape and diffusion from nearby PWNe. 

Like the Guitar, this is a multi-bubble bow shock, but here we see interior X-rays from the shocked pulsar wind. For the Guitar, \citet{Cordes1993} speculated that the bubbles were caused by ISM density irregularities or episodic variations in the PSR ${\dot E}$. With several H$\alpha$ bow shocks showing axisymmetric modulation, attribution to random ISM variations now seems implausible. \citet{Kerkwijk2008} proposed that the post-bow shock back-flow makes a collimated trail which feeds a bubble at its base until it has lengthened to $\delta r \approx 20-60 \times r_0$ (with $r_0$ the apex standoff distance) at  which time instabilities pinch off the flow. Thus bubbles spaced by $\delta r$ expand with a Sedov-Taylor solution for energy injection ${\dot E} \delta r/v_{PSR}$.

\citet{Morlino2015} propose another scenario in which the shock at the H$\alpha$ apex ionizes and accelerates only a fraction of the ISM gas; most of the H stays neutral, passing through this zone. Downstream, behind the Mach disk, the remaining gas is photo-ionized while embedded in the shocked pulsar wind. The resulting mass loading can cause the flow to decelerate and heat, with a resulting secondary shock flare. This may brighten the H$\alpha$ limb, adding `shoulders' to the H$\alpha$ bow shock. While this has some appeal as an explanation for the bubble 1 to bubble 2 transition in J2030, it does not naturally produce closed bubbles or additional downstream features.

We do not see the feeding axial flow of the \citet{Kerkwijk2008} picture, but perhaps the X-ray bubbles have already separated and cooled. Our structure does seem more likely to represent variable pulsar injection on the $\sim 300$\,yr crossing time. Some precedent comes from radio pulsars with episodic ${\dot P}$ (and hence ${\dot E}$) changes. J1841$-$0500 apparently changes ${\dot E}$ by $2.5\times$, with modes lasting over 2\,yr \citep{Camilo2012}. Perhaps the cylindrical PWN width represents the stalling radius for the high ${\dot E}$ state, while the spiral pinching represent low-power interludes. In any event, the apparent helical X-ray interior morphology should be verified with more sensitive imaging, but will likely prove a challenge to PWN shock modelers.

\acknowledgments
R.W.R. was supported in part by NASA grants G08-19049X and 80NSSC17K0024. 

\vspace{5mm}
\facilities{CXO, Gemini - GMOS}

\bigskip
\bibliographystyle{aasjournal}
\bibliography{J2030}


\end{document}